\documentclass{epl}

\usepackage{graphicx}

\begin{document}

\title{Spontaneous emergence of contrarian-like behaviour in an opinion
spreading model} \shorttitle{Spontaneous contrarian behaviour}

\author{Marta S.\ de La Lama\inst{1,2} \and Juan M.\ L\'opez\inst{1} \and Horacio
S.\ Wio\inst{1,3}}

\institute{ \inst{1} Instituto de F\'{\i}sica de
Cantabria (IFCA), CSIC--UC, E-39005 Santander, Spain\\
\inst{2} Departamento de F{\'\i}sica Moderna, Universidad de
Cantabria, Avda. Los Castros, E-39005 Santander, Spain\\
\inst{3} Centro Atomico Bariloche and Instituto Balseiro, 8400 San
Carlos de Bariloche, Argentina}

 \pacs{-0.50.+q}{Lattice
theory and statistics} 

\maketitle

\begin{abstract}
We introduce stochastic driving in the Sznajd model of opinion
spreading. This stochastic effect is meant to mimic a social
temperature, so that agents can take random decisions with a varying
probability. We show that a stochastic driving has a tremendous
impact on the system dynamics as a whole by inducing an
order-disorder nonequilibrium phase transition. Interestingly, under
certain conditions, this stochastic dynamics can spontaneously lead
to agents in the system who are analogous to Galam's {\it
contarians}.
\end{abstract}

\section{Introduction}

The study of complex systems, in particular the application of
statistical physics methods to social phenomena, has recently
attracted the attention of theoretical physicists
\cite{weidlich1,weidlich2,stauffer00,stauffer00p,Galam000}. From a
statistical physics perspective, social systems are modeled as a
collection of agents interacting through simple rules. 
In particular, the building (or the lack) of consensus in social
systems has attracted much attention in recent years. A number of
models have been considered in order to mimic the dynamics of
consensus in opinion formation, cultural dynamics, etc
\cite{nos+schn}. Among those models, Sznajd dynamics of opinion
spreading has been subject of a great deal of work in recent years.
Sznajd model is duly based on the trade union lema: "united we
stand, divided we fall", and has been studied on different network
topologies and for (slight) variations of the dynamics
\cite{sznajd01,sznajd02,stauffer1,stauffer2,stauffer3,nos+schn}.

An important aspect of social and economic systems, recently
discussed within opinion formation models, has been the presence of
some agents called {\it contrarians}-- namely, people who are in a
"nonconformist opposition". That is, people who always adopt the
opposite opinion to the majority
\cite{contrarians-00,contrarians-01,contrarians-02}. In stock
markets for instance, contrarians are those investors who buy shares
of stock when most others are selling and sell when others are
buying. The existence of a high proportion of contrarians in a
society may play an important r\^{o}le in social dynamics (think for
instance of referendums or stock market dynamics)
\cite{contrarians-00}.

In an attempt to include the contrarian effect in existing social
models, a number of previous studies have considered contrarian
agents as a initial condition, {\it i.e.} a given density of
contrarians is introduced in the model by hand for instance
\cite{contrarians-00,contrarians-01,contrarians-02}. This is
somewhat artificial and one would expect that simple models of
opinion spreading should spontaneously lead to the existence of a
fraction of contrarians among the population as some sort of
emergent property.

In this Letter we show that a contrarian effect can spontaneously
emerge when stochastic driving is included in the model. As a
typical model example we introduce stochastic dynamics on the Sznajd
model. This randomness in the update of an agent opinion is meant to
be a highly simplified description of the interplay between
fashion/propaganda and a collective climate parameter, which is
usually referred to as {\em social temperature} of the system
\cite{jml,bab-kup,tessotor}. We show here that social temperature in
Snajd-type models leads to the spontaneous appearance of contrarian
agents in the system.

\section{Mean-field approach}

Following the mean-field approach in \cite{sznajd02}, we have
considered the Sznajd model (the so-called ``two over one" case in
\cite{sznajd02}), where two agents are chosen randomly and, if they
are in consensus, then another randomly chosen agent is convinced by
them. The Fokker-Planck equation (FPE) for the probability
$\mathcal{P}(m,t)$ of having a ``magnetization" $m$, at time $t$
given a certain initial condition at time $t_0<t$ is given by
\begin{eqnarray}\label{eq:1}
\frac{\partial}{\partial t} \mathcal{P}(m,t) = &-& \frac{1}{2N}
\frac{\partial}{\partial m} \Bigl( m(1-m^2) \mathcal{P}(m,t) \Bigr)
\nonumber \\
& & \,\,\,\,\,\,\,\, + \left( \frac{1}{2N} \right)^2 \frac{\partial
^2} {\partial m^2} \Bigl( (1 -m^2) \mathcal{P}(m,t) \Bigr) +
O(N^{-3}),
\end{eqnarray}
where $N$ is the total number of agents. The magnetization density
$m = (N^+ - N^-)/N$ measures the opinion state of the system and
$N^+$, $N^-$ are the number of agents supporting the $+$ or the $-$
opinion, respectively (with $N^+ + N^- = N$).

We now include a {\em social temperature} effect in the model by
allowing the possibility that at every time step an agent follows
the rules of the Sznajd model with probability $p$ ($p \leq 1$),
while there is a probability $1-p$ that those rules are not
fulfilled (an agent adopts the opposite option than the one dictated
by the rules). Then, for a given probability $p$, we arrive at a FPE
for $\mathcal{P}_{p}(m,t)$ that reads
\begin{eqnarray}\label{eq:2}
\frac{\partial}{\partial t} \mathcal{P}_{p}(m,t) = &-& \frac{1}{2N}
\frac{\partial}{\partial m} \Bigl( \Bigl[ (6p-5)m - (2p-1) m^3)
\Bigr] \mathcal{P}_{p}(m,t) \Bigr)
\nonumber \\
& & \,\,\,\,\,\,\,\, + \left(\frac{1}{2N} \right)^2 \frac{\partial
^2} {\partial m^2} \Bigl( \Bigl[ 3-2p - (2p-1) m^2 \Bigr]
\mathcal{P}_{p}(m,t) \Bigr).
\end{eqnarray}
The stationary solution $\mathcal{P}_{p}^{stat}(m)$ results to be
\begin{eqnarray}\label{eq:3}
\mathcal{P}_{p}^{stat}(m) \approx \exp \left\{ N \int_{-1}^{m}
\frac{(6p-5) u (2p-1) u^3}{3-2p -(2p-1) u^2}\, du - \ln \Bigl[
N(3-2p -(2p-1) m^2) \Bigr] \right\}.
\end{eqnarray}
The analysis of this stationary solution for varying $p$ shows that
there is a threshold value, $p= p_c$, such that for $p>p_c$ the
system is {\em bistable} with a probability density
$\mathcal{P}_{p}^{stat}(m)$ having two maxima at $m_{\pm} = \pm
\sqrt{(6p-5)/(2p-1)}$. In this case the system gets ordered by
spontaneously selecting one of the stable solutions $m_\pm$. On the
contrary, for $p<p_c$ the system becomes {\em monostable} and
disordered with a magnetization density peaked at $m=0$ in which no
dominant opinion survives. The threshold $p_c$ can be calculated in
this mean-field approximation equating $m_{+} = m_{-}$, at which all
three extreme coalesce into a single minimum at $m=0$, so that we
find $p_c = 5/6$. This behavior is shown in Fig.\ 1.

\begin{figure}
\onefigure{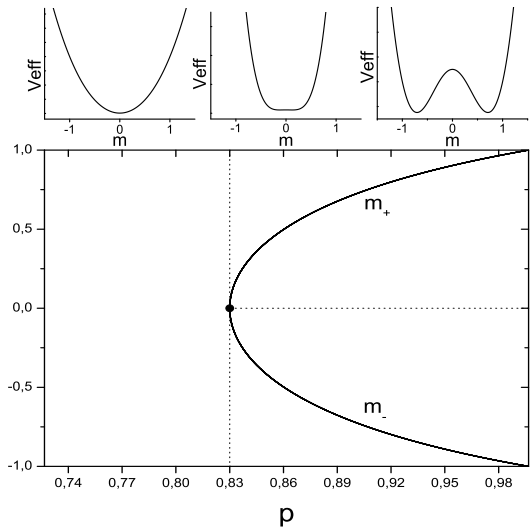} \caption{The effective mean-field potential
obtained from Eq. (\ref{eq:3}). For large values of p two minima
$m_\pm$ appear. These move continuously toward each other as the
threshold value $p_c=5/6$ is approached indicating a continuous
phase transition. {\it Bottom}: Position of the potential minima vs
p. When both extreme coalesce into $m=0$, the transition point is
reached. {\it Top}: Effective potential for (from left to right) $p
= 0.7$, $p=p_c=5/6$, and $p=0.9$.} \label{f.1}
\end{figure}

The picture emerging from the mean-field approach is clear. The
effect of including thermal fluctuations in Snajd type models
immediately leads to a contrarian-like effect. Some agents randomly
take decisions that oppose the rules of the model, indicating some
undecidedness in a fraction of the population. If such a fraction
overcomes the critical threshold ($1-p_c = 1/6 \approx 0.1667$) the
system will reach a stalemate situation, analogous to the {\it
contrarians} effect discussed in
\cite{contrarians-00,contrarians-01,contrarians-02}.

\section{Monte Carlo simulations}

In what follows we report on Monte Carlo simulations, in order to
test the above discussed mean-field results. We have studied the
model on regular lattices and small-world networks (which in the
limit of high rewiring probability should reproduce the mean-field
results). To make such an analysis, and in order to avoid the
spurious antiferromagnetic solution of the Sznajd original model, we
have studied a convenient variation proposed in \cite{sanchez}:
\begin{itemize}
\item rule $1'$: Chose an agent at random, say $i$, and if
$s_i \times s_{i+1} = 1,$ then $s_{i-1}$ and
$s_{i+2}$ adopt the direction of the selected pair $[i,i+1],$
\item rule $2'$: if $s_i \times s_{i+1} = -1,$ then $s_{i}$ adopts
the direction of $s_{i-1}$ and $s_{i+1}$ the direction of $s_{i+2}$.
\end{itemize}
Following \cite{sanchez}, in case of disagreement of the pair $(s_i,
s_{i+1})$, the rule $2'$ makes that the agent $i$ "feels more
comfortable" since it ends up with at least one neighbor having his
same opinion. This variation of Sznajd model does not affect the
basic behaviour and indeed has been shown to exhibit the same type
of scaling features as the original model.

We now introduce a stochastic mechanism in the dynamics as follows.
At each Monte Carlo step we assume that, with a probability $p$, the
rules are fully applied as indicated above, while the opposite
option to the one dictated by the rules happens with a probability
$1-p$. The probability $p$, in analogy with Weidlich
\cite{weidlich1,weidlich2} and Babinec \cite{bab-kup}, is defined
according to
\begin{equation}\label{sz-01}
p = \Lambda \exp \left[ \frac{\alpha}{\theta} \right],
\end{equation}
where $\alpha$ is some fixed parameter related to the strength of
nearest neighbour interactions (we assume $\alpha
> 0$), and $\theta$ is the collective climate parameter and plays the
r{\^o}le of a (social) temperature. The normalization constant is
$\Lambda^{-1} = \exp(\alpha/\theta) + \exp(-\alpha/\theta)$.

The asymptotic behaviour of $p$ is:
\begin{itemize}
\item if $\theta \rightarrow 0$, we have $p \rightarrow 1$,
indicating that without thermal fluctuations we recover S\'anchez
(and Sznajd) dynamics;

\item if $\theta \rightarrow \infty$, we have
$p \rightarrow 0.5$, the probability of fulfilling the model rules
or its opposite are the same. The model has a complete random
behavior.
\end{itemize}

Firstly we report on our results on the one-dimensional lattice.
Each lattice site is occupied by one agent with opinion (spin) $s_i
\in \{+1,-1\}$. Starting from a random initial condition we let the
system evolve towards its stationary state. For $\theta=0$ a
consensus state arises ($m_\pm =\pm 1$). However, as $\theta$ is
increased we observe transition towards the stalemate state as
predicted by mean-field theory, nonetheless the transition is
discontinuous (first order) as shown in Fig.\ 2.

\begin{figure}
\onefigure{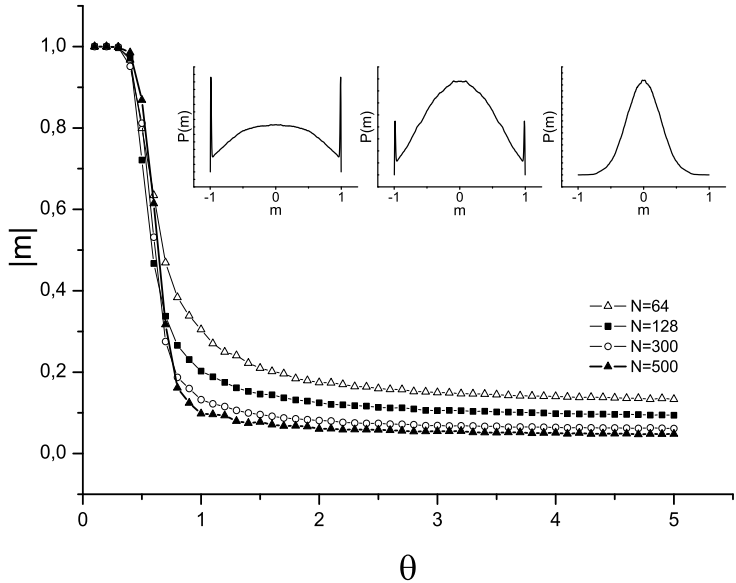} \caption{Discontinuous phase transition for the
one-dimensional lattice: the jump of the order parameter {\it m}
from $\pm 1$ to zero occurs abruptly. The discontinuity is more
evident for larger lattices. {\it Top}: Stationary distribution of
$m$ for $\theta = 0.44, 0.46$, and $0.6$ (from left to right), and
system size $N = 512$. For small values of $\theta$ a local maximum
starts to develop at $m=0$ and becomes the global maximum for
$\theta>\theta ^{\ast}$. Data were averaged over $5\times 10^3$
independent realizations.}\label{f.2}
\end{figure}

In order to compare with the mean-field results, we have studied our
model in a fully connected network, which is expected to behave as a
mean-field system. Indeed, in a fully connected network we observe
that the transition between order and stalemate state becomes
continuous. The qualitative agreement with mean-field results is
apparent in Fig.\ 3.

\begin{figure}
\onefigure{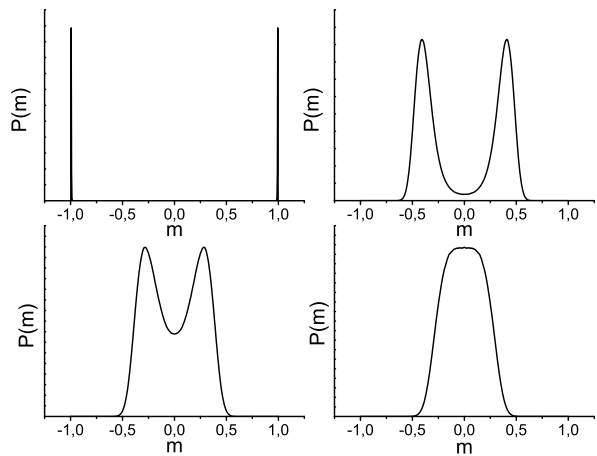} \caption{Stationary distribution of $m$ for a
fully-connected lattice of $N = 512$ agents. Data were averaged over
$100$ independent runs. From left to right and top to bottom $\theta
= 0.30, 2.30, 2.60$, and $2.90$. A second order transition towards a
``stalemate" state is apparent: the most probably values of the
order parameter {\it m} changes continuously from $m=\pm1$ ({\it
bistable}) to $m=0$ ({\it monostable}).} \label{f.3}
\end{figure}

We have also studied the intermediate cases between one-dimensional
and mean-field (infinite dimensional) limits by analyzing the model
on small-world networks. Starting from a regular a one-dimensional
lattice with periodic boundary conditions the links between
neighbours are rewired with a certain probability $r$ to a random
site. Even for small values of $r$ we observe a continuous phase
transition from order to disorder as predicted by mean-field theory.
We also observed that for increasing values of the rewiring
probability $r$, the critical temperature, $\theta ^{\ast}$, also
increases, as shown in Fig.4.

The critical density of contrarians that it is required to reach the
threshold ($\rho _c \sim 1-p(\theta ^{\ast})$) is relatively large
(the mean-field result $\rho _c \sim 0.33$ in the case 2 against 2).
However, such values are reasonable baring in mind that such a
density of contrarians corresponds to a statistical average,
dynamically generated by a large value of the social temperature
(that is, each agent could sustain, convince or change its opinion
dynamically). This mechanism is notably different from setting a
fixed number or density $\rho_c$ of agents in a ``nonconformist
opposition" that will {\em never} follow the rules, as was done in
previous studies
\cite{contrarians-00,contrarians-01,contrarians-02}. In physical
terms this difference is similar to the distinction between {\em
annealed} and {\em quenched} disorder.

\begin{figure}
\onefigure{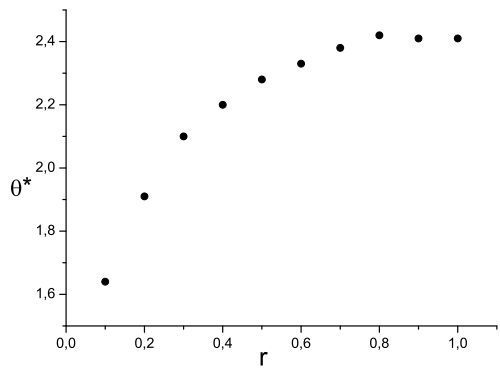}\caption{The critical temperature $\theta
^{\ast}$ {\it vs.} the rewiring probability $r$. All simulations
shown correspond to $N = 512$ agents and averages over $100$
independent realizations.} \label{f.4}
\end{figure}

\section{Conclusions}

We have proposed a dynamical mechanism that leads to a
contrarian-like effect analogous to the one described in
\cite{contrarians-00}. However, in contrast to
\cite{contrarians-00}, we found that contrarians may spontaneously
emerge from the dynamics when social temperature effects are taken
into account. For low temperatures the system gets to a consensus
where a majority opinion emerges just like in Sznajd type models.
However, when temperature is above a critical threshold the density
of contrarians is (on average) then high enough to make impossible
for the system to reach a consensus and the opinion is equally
divided between both options.

Here, we have considered different forms and the most convenient
prescriptions of the Sznajd model for the analytical and the
numerical analysis. However, we have checked that the phenomenon is
robust and does not depend on the particular form of the model.
Moreover, since the Sznajd model (as well as many other two state
opinion formation models) is similar to Ising type model
\cite{unify} up to a certain extend, we can regard our results as a
sophisticated manifestation in social systems of the ferromagnetic
transition in spin systems.

To conclude, the possibility of some external stochastic and/or
deterministic influence on the agents of an opinion formation model,
particularly regarding the possibility of some form of stochastic
resonance \cite{SR}, was recently analyzed by several authors
\cite{bab-kup,tessotor}. It would be of great interest to study such
an stochastic resonance effect (particularly its dependence on the
size of the system \cite{tessotor}) in our model. This can actualy
be done by including a fashion external field (for instance a
periodic signal) combined with the noise effect coming from the
social temperature. This is the subject of a forthcoming work.

\acknowledgments

We acknowledge financial support from Ministerio de Educaci\'on y
Ciencia (Spain) through Grant No. BFM2003-07749-C05-03 (Spain). MSL
is supported by a FPU fellowship (Spain). HSW thanks the European
Commission for the award of a {\it Marie Curie Chair} at Universidad
de Cantabria (Spain).

\end{document}